\documentclass[conference]{IEEEtran}
\usepackage{amsmath,amsfonts}
\usepackage{algorithmic}
\usepackage{algorithm}
\usepackage{array}
\usepackage[caption=false,font=normalsize,labelfont=sf,textfont=sf]{subfig}
\usepackage{textcomp}
\usepackage{stfloats}
\usepackage{url}
\usepackage{verbatim}
\usepackage{graphicx}
\usepackage{cite}

\begin{document}

\title{Grouped Pattern and Multi-Periodogram Algorithm for Range Estimation in ISAC Systems}

\author{\IEEEauthorblockN{Yi Geng}
	\IEEEauthorblockA{\textit{Cictmobile, China} \\
		gengyi@cictmobile.com}
                    \and
	\IEEEauthorblockN{Pan Cao}
	\IEEEauthorblockA{\textit{University of Hertfordshire, UK} \\
		p.cao@herts.ac.uk}
}

\maketitle

\begin{abstract}
This paper proposes a grouped pattern (GP) for sensing signals and a corresponding multi-periodogram algorithm for range estimation in integrated sensing and communications (ISAC) systems. GP partitions subcarriers into groups with an identical intra-group configuration replicated across groups, producing range profiles with periodic peaks and a structured multi-peak signature that improves low-SNR target detection. By identifying targets via cross-pattern peak validation, the proposed approach reduces missed detections and false alarms while requiring fewer dedicated sensing resources. Extensive simulations demonstrate a 16.5\% extended detection range and a 61\% reduced false alarm rate compared to conventional methods.
\end{abstract}


\section{Introduction}\label{sectionI}
Sensing signal (SS) pattern design is a central topic in integrated sensing and communications (ISAC) systems~\cite{10012421}. In low signal-noise ratio (SNR) sensing scenarios, sensing detection is often performed by applying constant false alarm rate (CFAR) processing to a sensing profile. However, CFAR faces an inherent tradeoff at low SNR: relaxing the detection threshold, i.e., using a higher probability of false alarm ($P_\text{fa}$), improves weak-target detectability but increases false alarms, whereas tightening the threshold suppresses false alarms but may cause weak targets to be missed. This motivates SS pattern and processing designs that provide more reliable detection cues beyond amplitude-only decisions.

\begin{figure}[!t]
\centerline{\includegraphics[width=1\linewidth, height=10cm, keepaspectratio]{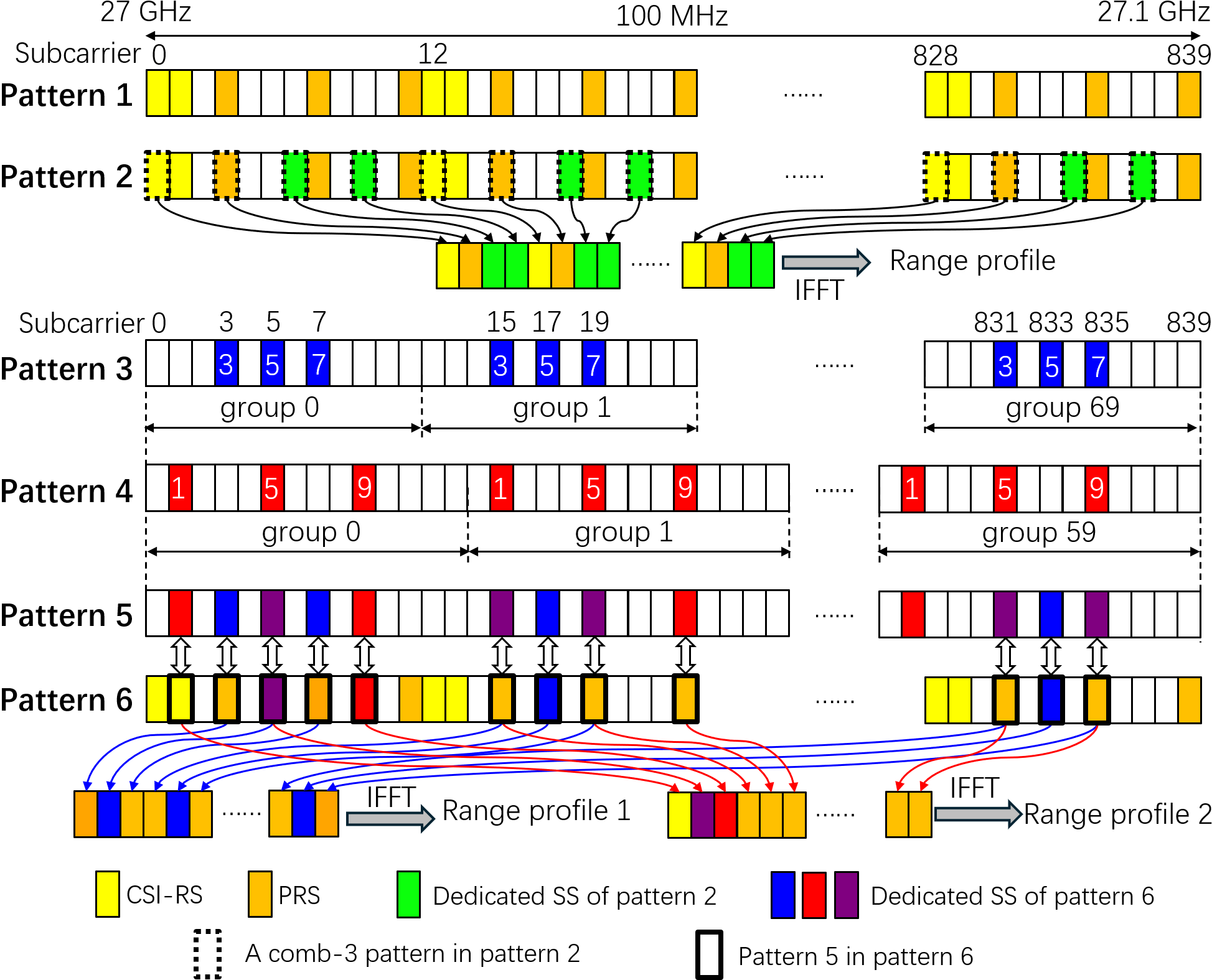}}
	\caption{Illustration of SS patterns used in this paper: RP (Pattern~2) and various GPs (Pattern 3-6) with their subcarrier distributions in the frequency domain.}
	\label{fig_1}
\end{figure}

Existing SS patterns can be categorized into two types: regular pattern (RP) and irregular pattern (IRP). 
RP refers to the equidistant distribution of SSs in the frequency domain, forming a comb-like pattern. There are three common ways to implement RP: (1) designing dedicated SSs~\cite{10908926}, which offers the best flexibility and performance but incurs high SS overhead. (2) Fully reusing existing 5G reference signals (RSs), where many 5G RSs such as positioning RS (PRS) and channel-state information RS (\mbox{CSI-RS}) can be reused as SSs~\cite{10833700},\cite{10694602}. However, most 5G RSs are sparsely distributed with RP in the frequency domain~\cite{zhang2023}. Superposing multiple RPs with different comb size rarely forms a denser RP~\cite{10788035}, limiting the feasibility of this approach. (3) A combined approach of 5G RSs and dedicated SSs, where 5G RSs are reused when their positions overlap with the desired SS pattern and dedicated SSs are transmitted otherwise. As Pattern~1 in Fig.~\ref{fig_1} shows the positions of CSI-RS and PRS within \mbox{27--27.1~GHz}, if a comb-3 SS pattern indicated by the dashed boxes in Pattern~2 is desired, dedicated SSs can be transmitted only on the subcarriers marked in green, while CSI-RS and PRS are reused as SSs on the other positions.

IRP, where SSs are non-equidistantly distributed, has emerged as a promising approach due to its flexibility in reusing 5G RSs and potential for enhanced sensing performance. Recent research in this area has predominantly focused on coprime-based patterns. Khosroshahi et al.~\cite{10901798} proposed combining PRS and PDSCH with different comb sizes to eliminate ghost targets in range estimation, effectively extending the unambiguous range. Similarly, Golzadeh et al.~\cite{10548650} introduced irregular PRS patterns to suppress ambiguity. Wang et al.~\cite{10626507} further developed frame-level sensing waveforms based on a double-coprime sampling structure to accommodate 5G frame constraints. While these approaches are valuable for ambiguity mitigation and RS reuse, robust detection in challenging low-SNR conditions remains difficult when decisions rely mainly on amplitude thresholding.

This paper proposes a \textit{grouped pattern} (\textit{GP}) and a \textit{multi-periodogram algorithm} for ISAC range estimation. We derive a factorized IFFT response for the resulting non-uniform samples, and resolve periodic-peak ambiguity via dual-GP cross-pattern validation. A structure-aware detector further improves low-SNR performance by reducing missed detections and false alarms compared with CFAR-based detection.

\section{Signal Models}\label{sectionII}
\subsection{Signal Model of RP}
RP is a fundamental SS allocation strategy in orthogonal frequency division multiplexing (OFDM)-based ISAC systems. Assuming SSs $d_{\text{TX}}(n)$ are transmitted on $N_\text{sc}$ subcarriers within a bandwidth $B$ to detect a target with range $R$. The received echo signal (ES) $d_{\text{RX}}(n)$ experiences a stochastic Rician channel model. A normalized ES $d(n)$ can be obtained by element-wise division between $d_{\text{RX}}(n)$ and $d_{\text{TX}}(n)$~\cite{5776640}:
\begin{equation}
	\label{eqn_1}
	d(n) = \frac{d_{\text{RX}}(n)}{d_{\text{TX}}(n)} = e^{-j2\pi\frac{Rn}{\Delta R N_\text{sc}}}, \quad n = 0, 1, \cdots, N_\text{sc}-1.
\end{equation}
where $\Delta R = \frac{\text{c}_0}{2B}$ denotes the range resolution, $\text{c}_0$ is the speed of light, $n$ denotes the subcarrier index.

To reduce SS overhead, SSs are transmitted in a RP with comb $C_\text{f}$~\cite{comb}. The resulting normalized ES of RP is
\begin{equation}
	\label{eqn_2}
	d_\text{RP}(m)  = e^{-j2\pi\frac{RC_\text{f}m}{\Delta R N_\text{sc}}}, \quad m = 0, 1, \cdots, \lfloor\frac{N_\text{sc}}{C_\text{f}}\rfloor-1.
\end{equation}
where $\left\lfloor \cdot \right\rfloor$ denotes the floor function. Applying an $\lfloor\frac{N_\text{sc}}{C_\text{f}}\rfloor$-point IFFT to $d_\text{RP}(m)$ yields $D_\text{RP}(k)$:
\begin{equation}\label{eqn_3}
	D_\text{RP}(k) = \sum_{m=0}^{\lfloor\frac{N_\text{sc}}{C_\text{f}}\rfloor-1} e^{-j2\pi\frac{RC_\text{f}m}{\Delta RN_\text{sc}}}e^{j2\pi\frac{mk}{N_\text{sc}/C_\text{f}}},k = 0,\cdots, \lfloor\frac{ N_\text{sc}}{C_\text{f}}\rfloor-1.
\end{equation}

The corresponding range profile is given by $|D_\text{RP}(k)|^2$. Then the range $R$ can be estimated as $\Delta R\cdot k_\text{p}$, where $k_\text{p}$ is the peak bin in the range profile. This approach, known as the periodogram algorithm, is the most common range estimation method in OFDM-based ISAC systems.

\subsection{Signal Model of Proposed Group Pattern}
While the RP reduces SS overhead through a comb pattern, it limits flexibility in SS allocation. To overcome this constraint, we propose an IRP named GP. GP divides $N_\text{sc}$ consecutive subcarriers into $N_\text{g}$ equal-sized groups, with each group containing $\lfloor\frac{N_\text{sc}}{N_\text{g}}\rfloor$ subcarriers. The  GP allows arbitrary number and pattern of SSs within each group, as long as the SS pattern remains identical across all groups. Within each group, $N_\text{s}$ SSs are transmitted at subcarrier positions
\[
s_0,\; s_1,\;\dots,\;s_i,\;\dots,\;s_{N_\text{s}-1},
\]
where $s_i$ is the intra-group subcarrier index of $i$-th SS within the group. Patterns~3 and 4 in Fig.~\ref{fig_1} illustrate two GP configurations:
\begin{itemize}
\item{Pattern~3: $N_\text{g,1} = 70$, $N_\text{s} = 3$ with $s_0=3$, \mbox{$s_1=5$}, $s_2=7$}
\item{Pattern~4: $N_\text{g,2} = 60$, $N_\text{s} = 3$ with $s_0=1$, \mbox{$s_1=5$}, $s_2=9$}
\end{itemize}

Among the $N_\text{g}N_\text{s}$ SSs indexed by $m$, SSs with indices $0,\ldots,N_\text{s}-1$ belong to group~0, SSs with indices $N_\text{s},\ldots,2N_\text{s}-1$ belong to group~1, and so on. We define $l(m)$ as the intra-group SS index function:
\begin{equation}\label{eqn_4}
  l(m) = 
  \begin{cases} 
    s_0, & \text{if } m \bmod N_\text{s} = 0,\\
    s_1, & \text{if } m \bmod N_\text{s} = 1,\\
    ~\vdots & \\
    s_{N_\text{s}-1}, & \text{if } m \bmod N_\text{s} = N_\text{s}-1.
  \end{cases}
\end{equation}

The GP yields $N_\text{g}N_\text{s}$ ESs that are non-uniformly sampled over the $N_\text{sc}$ subcarriers. A key step is to concatenate these ESs into a sequential signal $d_\text{GP}(m)$ indexed by $m$, despite their irregular frequency positions:
\begin{equation}\label{eqn_5}
d_\text{GP}(m) = e^{-j2\pi \frac{R}{\Delta R N_\text{sc}}(\frac{N_\text{sc}}{N_\text{g}}\lfloor \frac{m}{N_\text{s}} \rfloor + l(m))},\\m = 0, 1, \cdots, N_\text{g}N_\text{s}-1.
\end{equation}

In \eqref{eqn_5}, $\lfloor \frac{m}{N_\text{s}} \rfloor$ identifies the group index for the $m$-th signal. Since SSs are not transmitted on all subcarriers, $d_\text{GP}(m)$ represents a subset of $d(n)$ in \eqref{eqn_1}. Taking Pattern~3 in Fig.~\ref{fig_1} as an example, the intra-group SS index becomes
\begin{equation}\label{eqn_6}
	l(m) = 
	\begin{cases} 
		3, & \text{if}~m \bmod 3 = 0,\\
        5, & \text{if}~m \bmod 3 = 1,\\
        7, & \text{if}~m \bmod 3 = 2,
	\end{cases}
\end{equation}
where $N_\text{g}=70$, $N_\text{s}=3$, and $N_\text{sc}=840$. Combining \eqref{eqn_1} and \eqref{eqn_5}, the relationship between the index $m$ and the original subcarrier index $n$ can be written as
\begin{equation}
	\label{eqn_7}
n = 12\big\lfloor \tfrac{m}{3}\big\rfloor + l(m),\\m = 0, 1, \cdots, 209.
\end{equation}

Therefore, $d_\text{GP}(m)$ consists of the following ES samples:
\begin{equation}
	\label{eqn_8}
\underbrace{d(3), d(5), d(7)}_{\text{group 0}},
\underbrace{d(15), d(17), d(19)}_{\text{group 1}},
\ldots,
\underbrace{d(831), d(833), d(835)}_{\text{group 69}}.
\end{equation}

\section{Proposed Multi-Periodogram Algorithm}\label{sectionIII}
\subsection{Principle of Multi-Periodogram Algorithm}
Although $d_\text{GP}(m)$ in \eqref{eqn_5} is a non-uniformly sampled sequence in the frequency domain, we treat it as an indexed sequence and apply an $N_\text{g}N_\text{s}$-point IFFT, yielding
\begin{multline}\label{eqn_9}
D_\text{GP}(k) = \sum_{m=0}^{N_\text{g}N_\text{s}-1} e^{-j2\pi \frac{R}{\Delta R N_\text{sc}}(\frac{N_\text{sc}}{N_\text{g}}\lfloor \frac{m}{N_\text{s}} \rfloor + l(m))}e^{j2\pi\frac{mk}{N_\text{g}N_\text{s}}}, \\k = 0, 1,\cdots, N_\text{g}N_\text{s}-1.
\end{multline}
where $k$ is the bin index of the range profile. 

Since all $N_\text{s}$ SSs within the same group share an identical group index, we define: (1) $p$ as the group index, ranging from $0$ to $N_\text{g}-1$; and (2) $q$ as the SS index within each group, ranging from $0$ to $N_\text{s}-1$. The index $m$ can then be written as
\begin{equation}\label{eqn_10}
m = pN_\text{s} + q, \quad
p = 0,\dots,N_\text{g}-1,\; q = 0,\dots,N_\text{s}-1.
\end{equation}
Accordingly,
\begin{equation}\label{eqn_11}
\lfloor \frac{m}{N_\text{s}} \rfloor = p,
\qquad
l(m) = l(q),
\end{equation}
then \eqref{eqn_9} can be rewritten as
\begin{multline}\label{eqn_12}
D_\text{GP}(k) = \sum_{p=0}^{N_\text{g}-1}\sum_{q=0}^{N_\text{s
}-1} e^{-j2\pi \frac{R}{\Delta R N_\text{sc}}[\frac{N_\text{sc}p}{N_\text{g}} + l(q)]}e^{j2\pi\frac{(pN_\text{s}+q)k}{N_\text{g}N_\text{s}}}, \\k = 0, 1,\cdots, N_\text{g}N_\text{s}-1.
\end{multline}

Separating the terms that depend on $p$ and $q$, we obtain
\begin{multline}\label{eqn_13}
D_\text{GP}(k)
\\= \underbrace{\Biggl(\sum_{p=0}^{N_\text{g}-1}
\overbrace{e^{-j2\pi \frac{Rp}{N_\text{g}\Delta R}}}^{\text{Term 1}}
\overbrace{e^{j2\pi\frac{pk}{N_\text{g}}}}^{\text{Term 2}}
\Biggr)}_{D_\text{GP}^{(1)}(k)}
\,
\underbrace{\Biggl(\sum_{q=0}^{N_\text{s}-1}
e^{-j2\pi \frac{R\, l(q)}{\Delta R N_\text{sc}}}
e^{j2\pi\frac{qk}{N_\text{g}N_\text{s}}}
\Biggr)}_{D_\text{GP}^{(2)}(k)},\\
k = 0, 1,\cdots, N_\text{g}N_\text{s}-1 .
\end{multline}

Equation \eqref{eqn_13} factors the $N_\text{g}N_\text{s}$-point IFFT into a product of two $k$-dependent terms, where $D_\text{GP}^{(1)}(k)$ collects the contributions across groups and $D_\text{GP}^{(2)}(k)$ collects the contributions within each group. The reformulation from \eqref{eqn_12} to \eqref{eqn_13} provides critical insights of how $N_\text{g}$, $N_\text{s}$, and SS configuration $l(q)$ collectively shape the range profile $|D_\text{GP}(k)|^2$. From \eqref{eqn_13}, the range profile $|D_\text{GP}(k)|^2$ can be expressed as the Hadamard product (pointwise product, denoted by $\odot$) of two components:
\begin{equation}\label{eqn_14}
	|D_\text{GP}(k)|^2 = |D_\text{GP}^{(1)}(k)|^2\odot |D_\text{GP}^{(2)}(k)|^2.
\end{equation}

Term~1 and Term~2 in \eqref{eqn_13} cancel each other when \mbox{$\frac{Rp}{N_\text{g}\Delta R} = \frac{pk}{N_\text{g}}$}, creating a peak at bin $\frac{R}{\Delta R}$ in $|D_\text{GP}^{(1)}(k)|^2$. Moreover, due to Term~2's periodicity of $N_\text{g}$, $|D_\text{GP}^{(1)}(k)|^2$ forms a periodic structure with identical-amplitude peaks spaced every $N_\text{g}$ bins. As a result, $|D_\text{GP}^{(1)}(k)|^2$ presents a periodic set of peaks located at
$k=\frac{R}{\Delta R}+tN_\text{g}$, where $t$ is an integer. Crucially, only the peak at bin $\frac{R}{\Delta R}$ contains genuine range information, while other peaks introduce range ambiguities, which motivates the dual-GP cross-pattern validation described next. For the second component, $D_\text{GP}^{(2)}(k)$ can be interpreted as a superposition of $N_\text{s}$ complex exponentials, each represented as $E(q, k)$ with a unique frequency $f(q)$ and initial phase $\theta(q)$:
\begin{equation}\label{eqn_15}
	D_\text{GP}^{(2)}(k) = \sum_{q=0}^{N_{\text{s}}-1} E(q, k) = \sum_{q=0}^{N_{\text{s}}-1}e^{j[2\pi f(q)k-\theta(q)]},
\end{equation}
where
\begin{equation}\label{eqn_16}
	f(q) = \frac{q}{N_\text{g}N_\text{s}},~\theta(q) = \frac{2\pi R \cdot l(q)}{\Delta R N_\text{sc}},\quad q = 0, 1,\cdots, N_\text{s}-1.
\end{equation}

\textit{Remark:} The phase $\theta(q)$ of each $E(q, k)$ is directly governed by the GP configuration $l(q)$. This dependency causes $D_\text{GP}^{(2)}(k)$ to exhibit non-uniform amplitude distribution across different bins. This mechanism resembles beamforming techniques, where phase adjustments across antenna elements enable directed energy concentration. The Hadamard product in \eqref{eqn_14} performs bin-by-bin (e.g., $k$-by-$k$) amplitude multiplication between $|D_\text{GP}^{(1)}(k)|^2$ and $|D_\text{GP}^{(2)}(k)|^2$, sculpting the range profile. A fundamental distinction from periodogram algorithms, which yield a single peak per target, is that the proposed method deliberately generates a structured multi-peak pattern. This distinctive characteristic forms the basis for our nomenclature: the multi-periodogram algorithm.

\subsection{Ambiguity Mitigation via Dual-GP Cross-Pattern Validation}\label{sectionIII_ambiguity}
The periodic peaks in $|D_\text{GP}^{(1)}(k)|^2$ imply that a single GP generally produces multiple candidate peaks for one target, located at $k=\frac{R}{\Delta R}+tN_\text{g}$. To mitigate this ambiguity, we employ two GPs with different group numbers, denoted by $N_{\text{g,1}}$ and $N_{\text{g,2}}$. For a true target, the two resulting range profiles share a common peak at $k=\frac{R}{\Delta R}$, while the other periodic peaks occur at different bin locations due to the distinct periods $N_{\text{g,1}}$ and $N_{\text{g,2}}$. Therefore, the target range bin can be identified by cross-pattern peak validation, i.e., selecting the peak that appears consistently in both profiles.



\section{Simulation Results and Performance Analysis}\label{sectionIV}
We consider a monostatic sensing setup operating at 27~GHz with $B=100$~MHz and $N_{\mathrm{sc}}=840$. Target echoes are generated over a two-way free-space propagation channel with additive thermal noise. Targets follow the Swerling~1 model.
\subsection{Ambiguity Mitigation}\label{sectionIV_ambiguity}
\begin{figure}[t]
	\centering
	\subfloat[]{\label{fig_2a}\includegraphics[width=0.48\columnwidth]{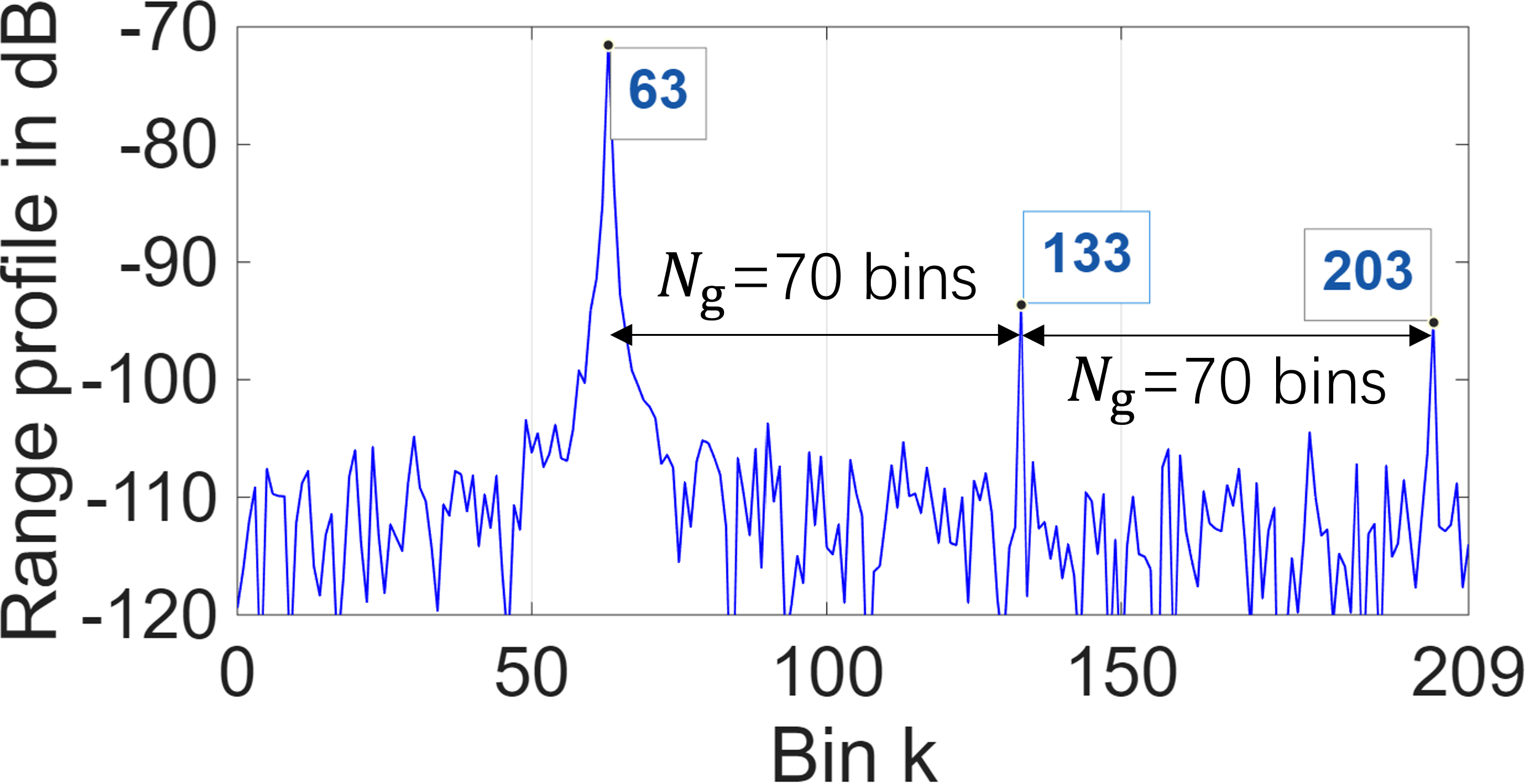}}\quad
   	\subfloat[]{\label{fig_2b}\includegraphics[width=0.48\columnwidth]{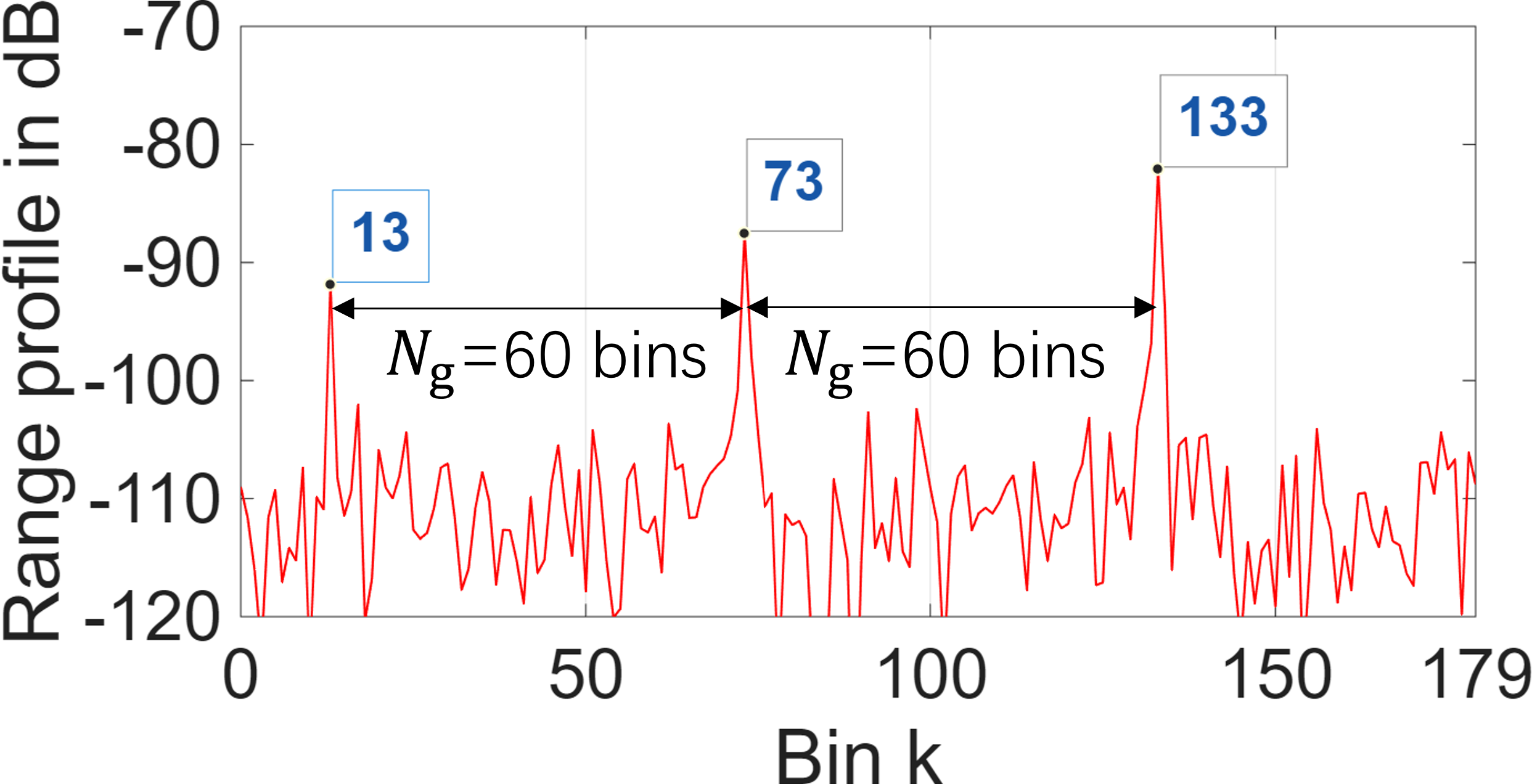}}\quad
	\subfloat[]{\label{fig_2c}\includegraphics[width=0.48\columnwidth]{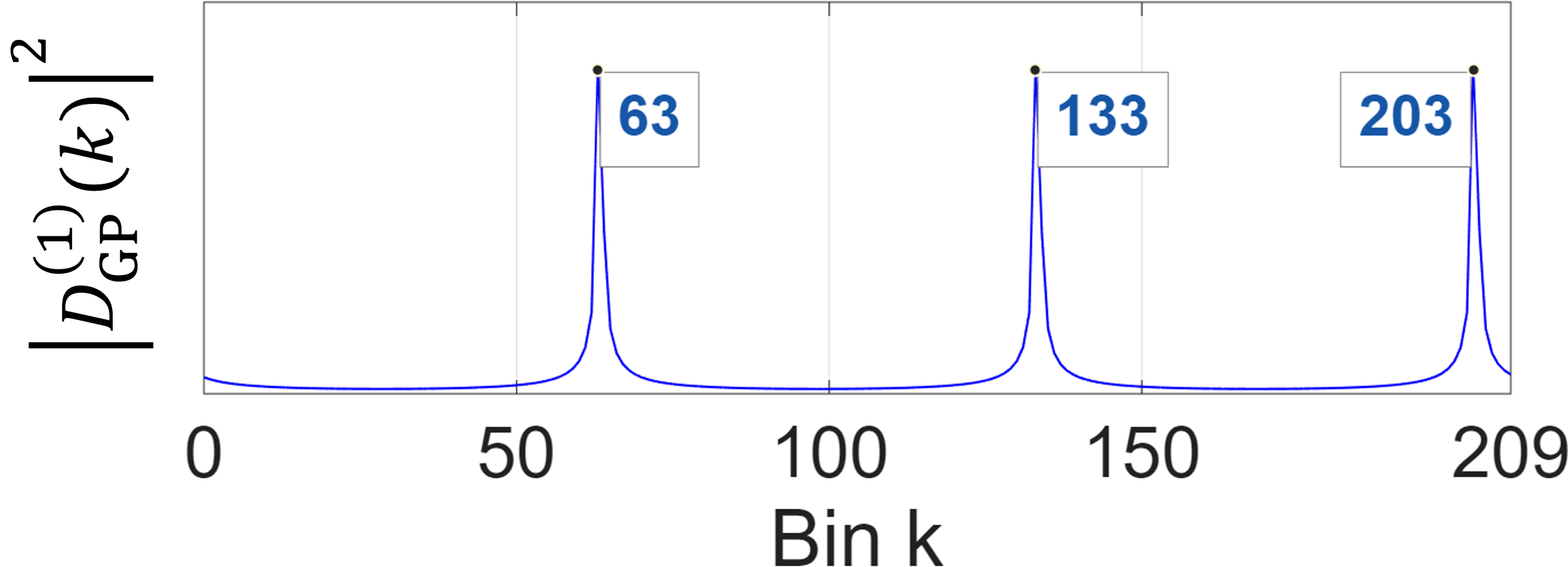}}\quad
   	\subfloat[]{\label{fig_2d}\includegraphics[width=0.48\columnwidth]{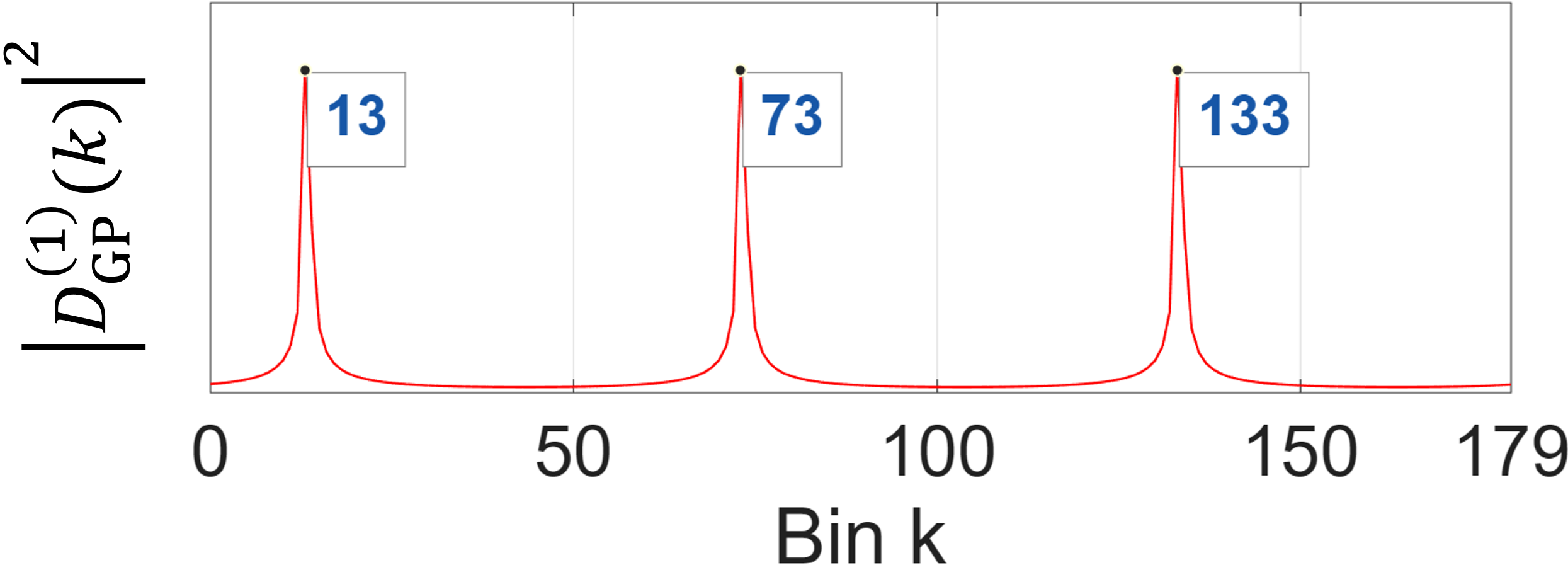}}\quad
	\subfloat[]{\label{fig_2e}\includegraphics[width=0.48\columnwidth]{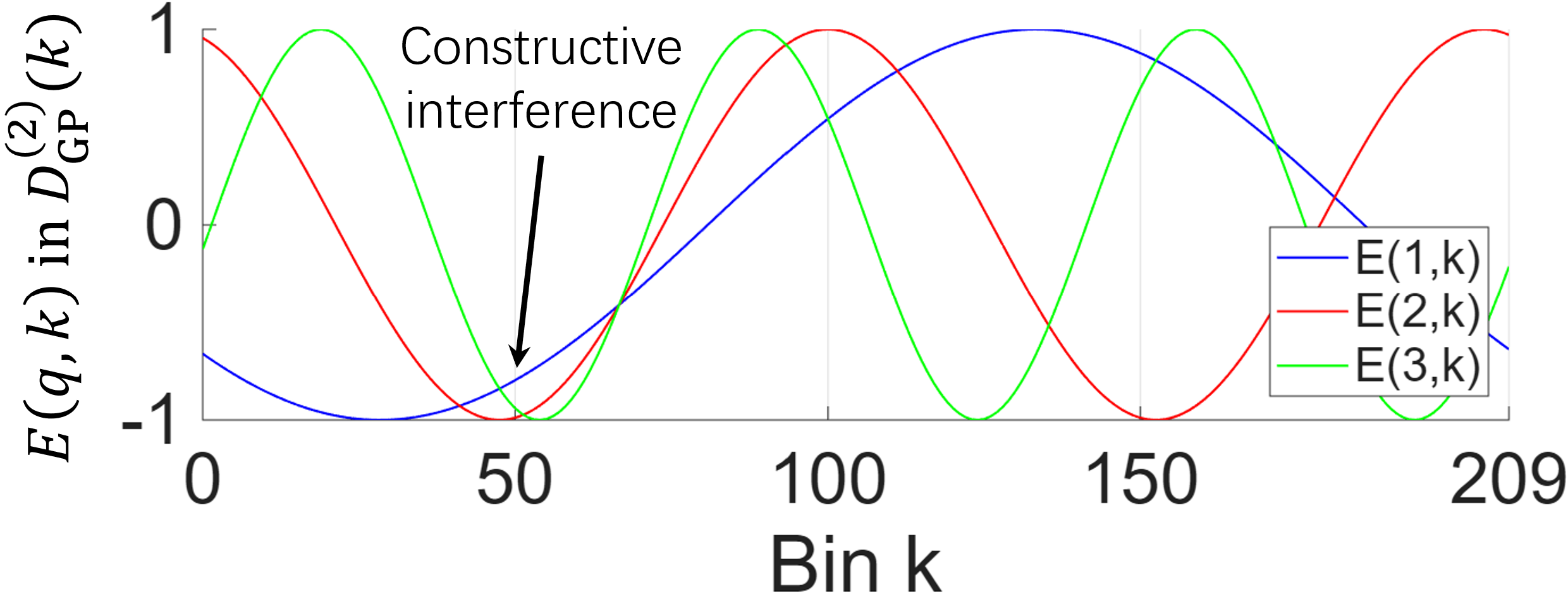}}\quad
   	\subfloat[]{\label{fig_2f}\includegraphics[width=0.48\columnwidth]{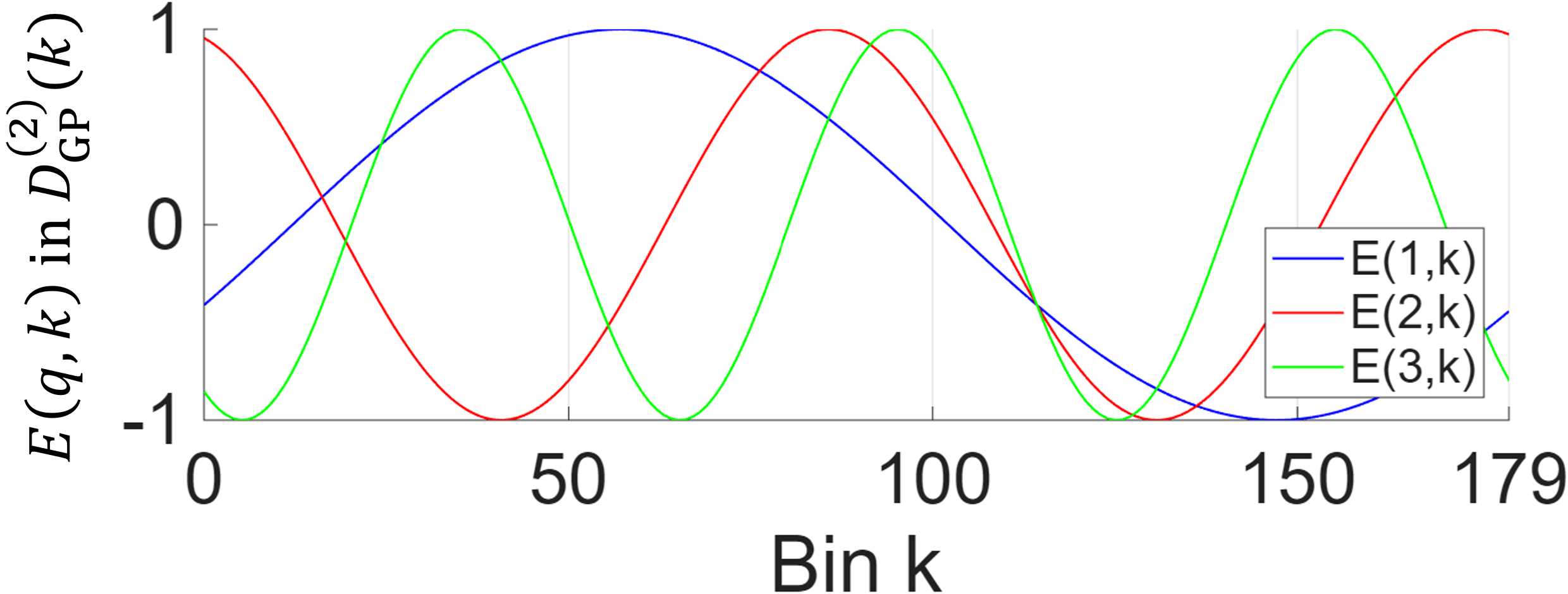}}\quad
	\subfloat[]{\label{fig_2g}\includegraphics[width=0.48\columnwidth]{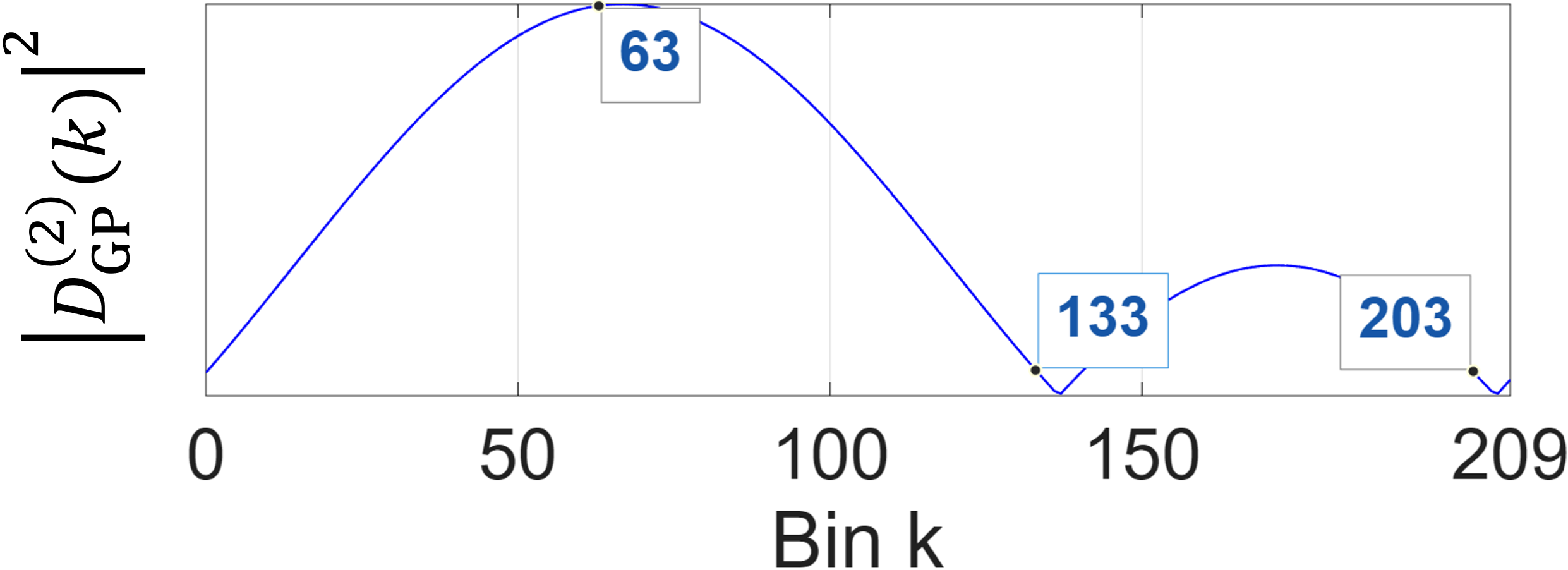}}\quad
    \subfloat[]{\label{fig_2h}\includegraphics[width=0.48\columnwidth]{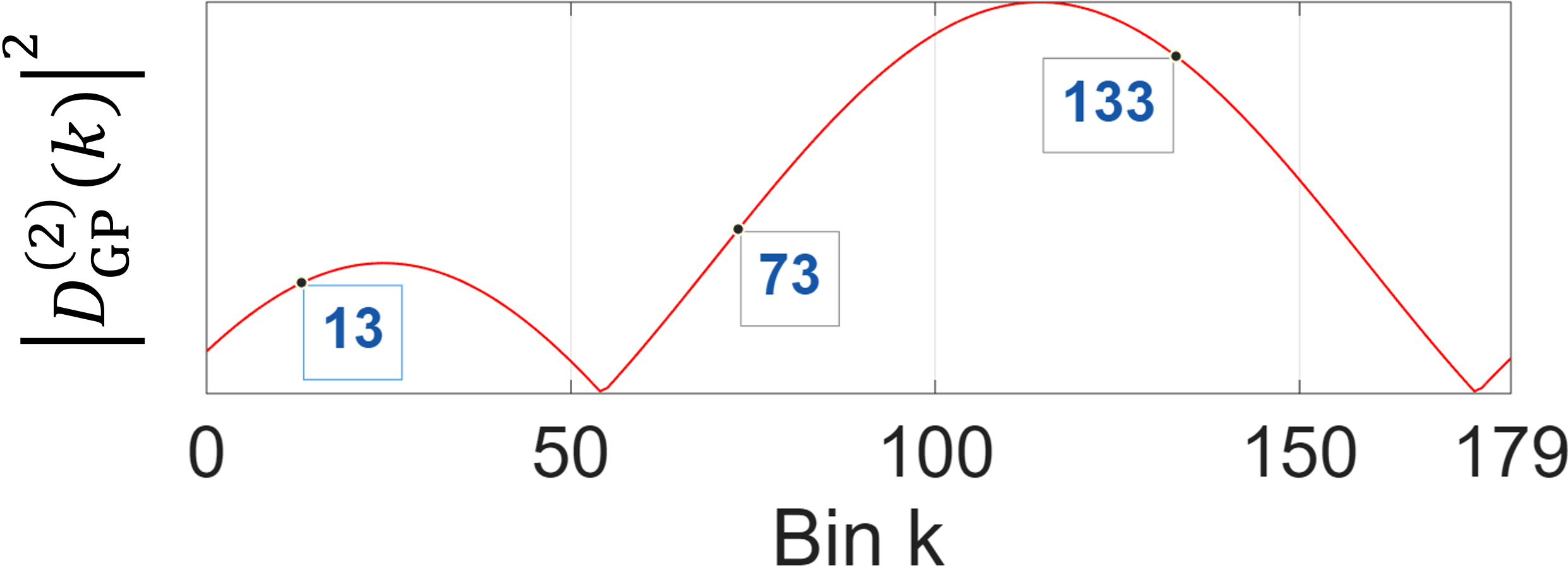}}\\
	\caption{Target detection at 200~m using GP: (a) range profile derived from Pattern~3, (b) range profile derived from Pattern~4, (c) $|D_\text{GP}^{(1)}(k)|^2$ of Pattern~3, (d) $|D_\text{GP}^{(1)}(k)|^2$ of Pattern~4, (e) $E(q, k)$ constituting $D_\text{GP}^{(2)}(k)$ of Pattern~3, (f) $E(q, k)$ constituting $D_\text{GP}^{(2)}(k)$ of Pattern~4, (g) $|D_\text{GP}^{(2)}(k)|^2$ of Pattern~3, (h) $|D_\text{GP}^{(2)}(k)|^2$ of Pattern~4.}
	\label{fig_2}
\end{figure}

We first illustrate the ambiguity-mitigation effect of the proposed dual-GP cross-pattern validation using Patterns~3 and 4 in Fig.~\ref{fig_1}. Patterns~3 and 4 partition the subcarriers into 70 and 60 groups, respectively. Their superposition yields Pattern~5, and Pattern~6 further integrates Pattern~5 with Pattern~1 to reduce dedicated SS overhead by reusing CSI-RS and PRS.
By extracting ESs corresponding to Patterns~3 and 4 (as depicted in the diagram below Pattern~6 in Fig.~\ref{fig_1}) and applying the proposed algorithm, two range profiles are generated as shown in Figs.~\ref{fig_2}(a) and \ref{fig_2}(b) for a target at 200~m. The periodic peaks exhibit intervals of 70 and 60~bins, respectively. Both profiles reveal a common peak at bin~133, confirming the target at 199.5~m ($133\Delta R$, where \mbox{$\Delta R$ = 1.5~m}).

\subsection{Range Profile Analysis}
Figs.~\ref{fig_2}(c) and \ref{fig_2}(d) illustrate $|D_\text{GP}^{(1)}(k)|^2$ components that contribute to the range profiles depicted in Figs.~\ref{fig_2}(a) and~\ref{fig_2}(b), respectively. Both components exhibit periodic peak structures with a common peak at bin~133. The exponentials $E(q, k)$ that collectively constitute $D_\text{GP}^{(2)}(k)$ for Patterns~3 and 4 are illustrated in Figs.~\ref{fig_2}(e) and \ref{fig_2}(f). Subsequently, Figs.~\ref{fig_2}(g) and \ref{fig_2}(h) present the superpositions of these respective exponentials, demonstrating the amplitude distribution of $|D_\text{GP}^{(2)}(k)|^2$ across different bins for Patterns~3 and 4.

The Hadamard product in \eqref{eqn_14} represents a critical operation where the signals in Fig.~\ref{fig_2}(c) modulates the signals in Fig.~\ref{fig_2}(g). The three exponentials depicted in Fig.~\ref{fig_2}(e) reach their amplitude minima simultaneously near bin~63, creating powerful constructive interference. This phase alignment generates the pronounced amplitude enhancement visible in Fig.~\ref{fig_2}(g) at bin~63. Consequently, through the bin-wise multiplication of the Hadamard product, this localized amplification manifests as the dominant peak at bin~63 in the range profile shown in Fig.~\ref{fig_2}(a).

\begin{figure}[t]
	\centering
	\subfloat[]{\label{fig_3a}\includegraphics[width=0.48\columnwidth]{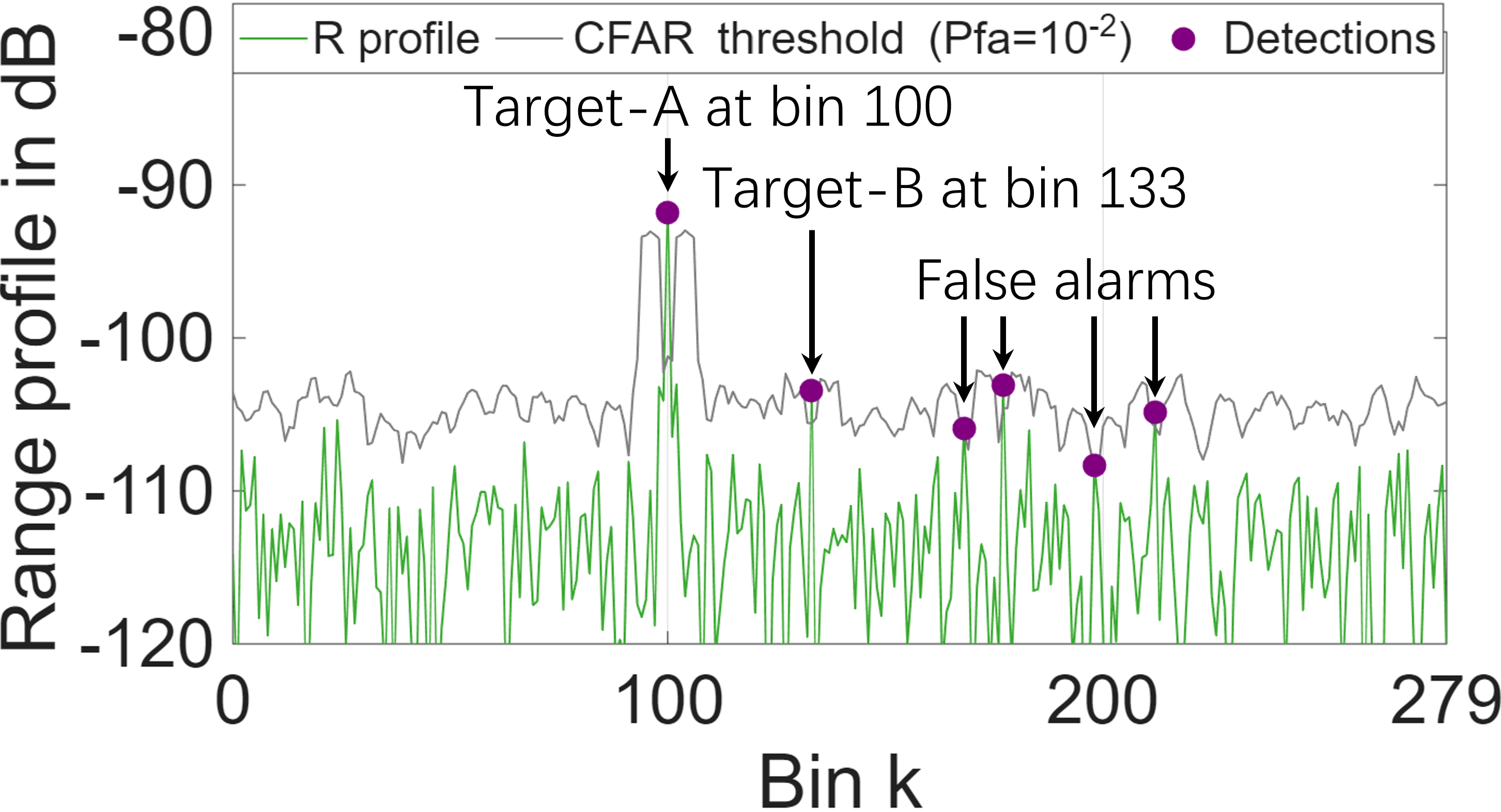}}\quad
   	   	\subfloat[]{\label{fig_3b}\includegraphics[width=0.48\columnwidth]{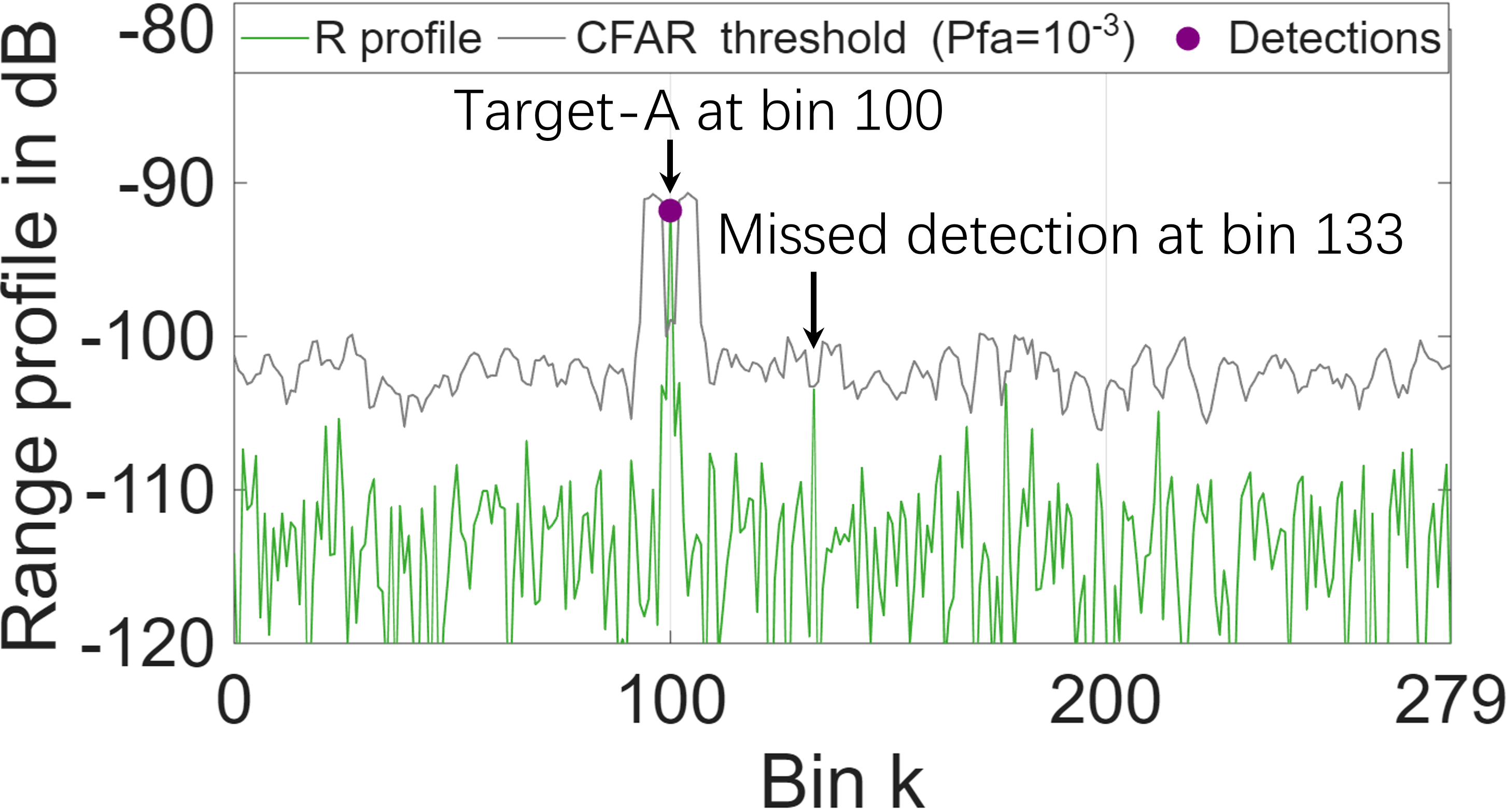}}\quad
        \subfloat[]{\label{fig_3c}\includegraphics[width=0.48\columnwidth]{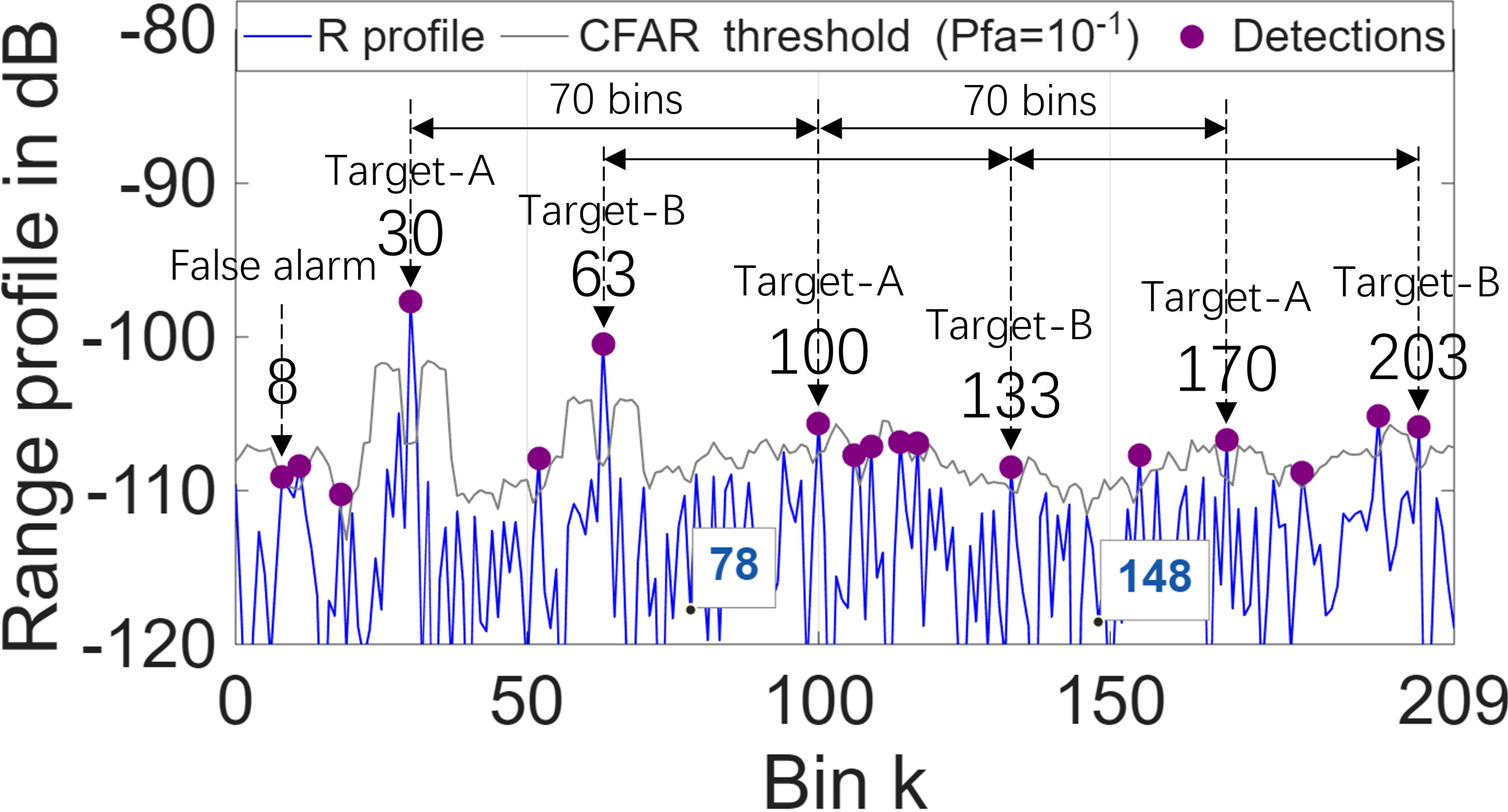}}\quad
        \subfloat[]{\label{fig_3d}\includegraphics[width=0.48\columnwidth]{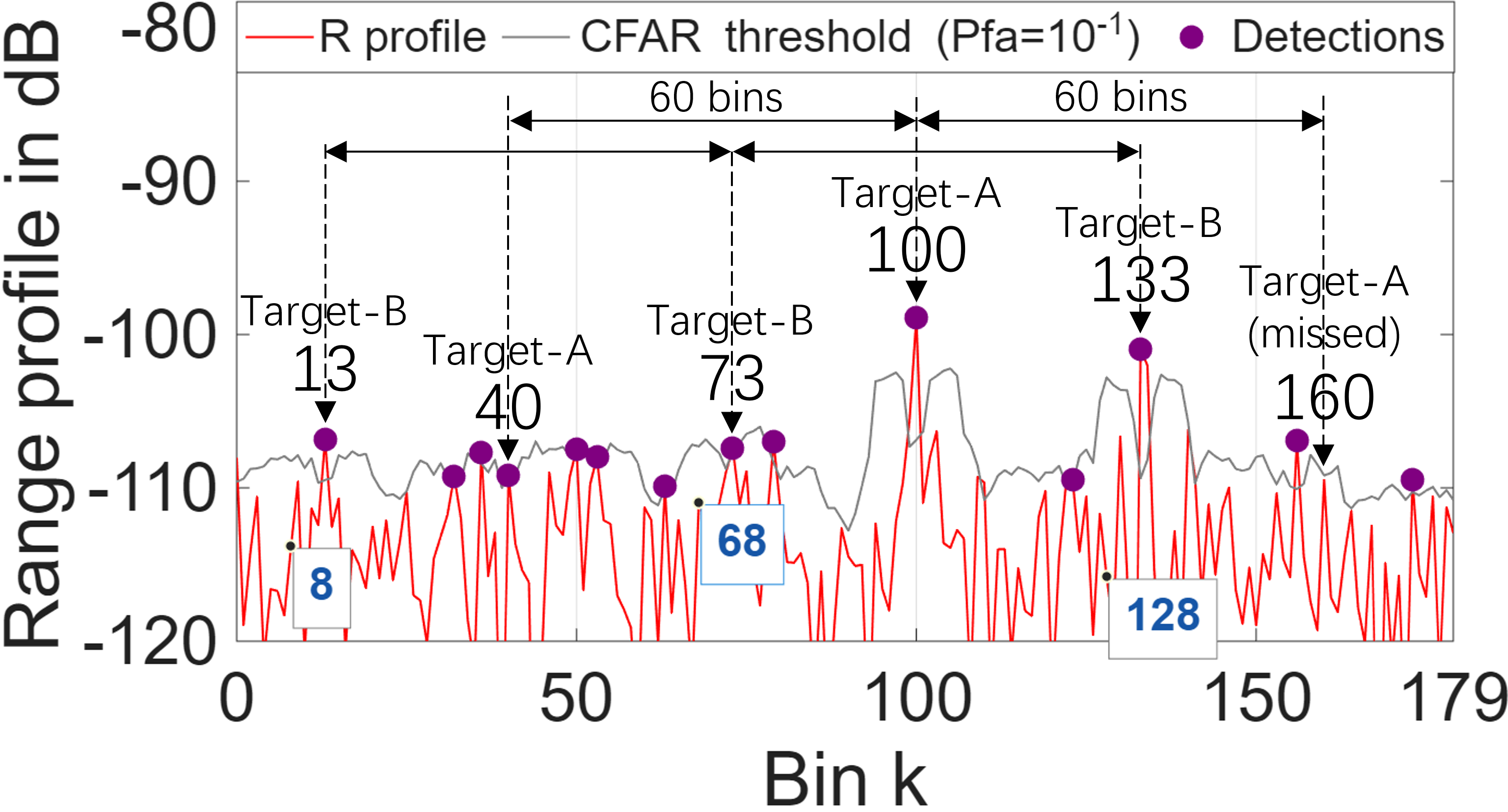}}\\
	\caption{Detecting two targets at 150~m and 200~m with CFAR and GNN: (a) Pattern~2 and $P_\text{fa}$ of $10^{-2}$, (b) Pattern~2 and $P_\text{fa}$ of $10^{-3}$, (c) Pattern~3 and $P_\text{fa}$ of $10^{-1}$, (d) Pattern~4 and $P_\text{fa}$ of $10^{-1}$.}
	\label{fig_3}
\end{figure}

\subsection{Superior Performance in Low SNR Scenarios}
The proposed algorithm demonstrates significant advantages in low-SNR environments where conventional methods generally underperform due to the trade-off between detection sensitivity and false alarm. Fig.~\ref{fig_3} presents a comparative analysis between the periodogram algorithm and the proposed approach when detecting two targets at 150~m (\mbox{target-A}) and 200~m (target-B). Fig.~\ref{fig_3}(a) and Fig.~\ref{fig_3}(b) illustrate the detections using Pattern~2, periodogram algorithm, and CFAR with $P_\text{fa}$ of $10^{-2}$ and $10^{-3}$, respectively. Target-B, which should produce a peak at bin~133, exemplifies the limitation of conventional methods: with $P_\text{fa}$ of $10^{-2}$, target-B is detected but accompanied by numerous false alarms, while reducing $P_\text{fa}$ to $10^{-3}$ to mitigate false alarms results in a missed detection. This is because the peak amplitude of target-B is close to the noise level, and the CFAR algorithm relies solely on peak amplitude for detection, making it unable to balance false alarms and missed detections.

In contrast, Figs.~\ref{fig_3}(c) and \ref{fig_3}(d) demonstrate the detections with a higher $P_\text{fa}$ of $10^{-1}$ using Patterns~3, 4 and the proposed algorithm. Due to this extremely high $P_\text{fa}$ setting, 5 out of 6 low-amplitude peaks generated by target-A are successfully detected: bins~30, 100, and 170 in Fig.~\ref{fig_3}(c), and bins 40 and 100 in Fig.~\ref{fig_3}(d), with only the peak at bin~160 in Fig.~\ref{fig_3}(d) being missed. Simultaneously, this high $P_\text{fa}$ setting inevitably produces numerous false alarms. The key advantage of our proposed algorithm in reducing both missed detections and false alarms lies in leveraging the inherent periodic structure of target-generated peaks—a characteristic that random noise-induced detections lack. In Fig.~\ref{fig_3}(c), detections at bins~30, 100, and 170 exhibit the expected 70-bin separation characteristic of Pattern~3. Similarly, in Fig.~\ref{fig_3}(d), detections at bins~40 and 100 follow the \mbox{60-bin} periodicity of Pattern~4 (though the peak at bin~160 falls below the threshold). Critically, both patterns share a common detection at bin~100, providing robust evidence of a target at~150 m ($100\Delta R$). This verification mechanism significantly enhances detection reliability, as noise-induced detections would rarely satisfy such strict constraints. The same verification mechanism effectively suppresses false alarms as well. Consider the detection at bin~8 in Fig.~\ref{fig_3}(c), if this detection represented a genuine target, it would generate corresponding periodic peaks at bins~78 and 148 in Fig.~\ref{fig_3}(c), as well as at bins~8, 68, and 128 in Fig.~\ref{fig_3}(d). The absence of detections at these expected bins allows the algorithm to confidently classify the detection at bin~8 as noise-induced, eliminating it from the false alarm count.

\subsection{Comprehensive Performance Evaluation}
\begin{table}[!t]
\caption{Comparison of SS overhead between Patterns 2 and 6\label{tab:table1}}
\centering
\begin{tabular}{|c|c|c|}
\hline
Parameter & Pattern 2 (RP) & Pattern 6 (GP)\\
\hline
Configuration & Pattern 1 + 2 & Patterns 1 + 3 + 4\\
\hline
Total sensing subcarriers & 280 & 300\\
\hline
Reusing CSI-RS/PRS & 142 & 200\\
\hline
Dedicated SS overhead & 138 (49.3\%) & 100 (33.3\%)\\
\hline
\end{tabular}
\end{table}
We conducted Monte Carlo simulations to compare the performance of the periodogram and multi-periodogram algorithms using Patterns~2 and~6, respectively. Table~\ref{tab:table1} summarizes the configurations of these patterns. Pattern~6 uses more sensing subcarriers than Pattern~2 (300 vs.\ 280), but it reuses more CSI-RS/PRS positions (200 vs.\ 142). As a result, Pattern~6 requires fewer dedicated SSs (100 vs.\ 138), thanks to the overlapping subcarriers between Patterns~3 and~4 (highlighted in purple in Fig.~\ref{fig_1}) that simultaneously serve both GPs.

The simulations detect a single target with ranges from 20 to 240~m in steps of 20~m. For each range, 500 independent Monte Carlo trials are performed, yielding 6000 simulation runs in total. The ES SNR in radar systems follows~\cite{5494616}:
\begin{equation}
	\label{eqn_17}
	SNR = 10\text{log}_{10}\frac{C_\text{PN}\cdot \sigma_\text{RCS}}{(4\pi)^3Bf_\text{c}^2R^4},
\end{equation}
where $C_\text{PN}$ is a constant determined by transmit power, antenna gain, noise power, and Boltzmann constant, and $\sigma_\text{RCS}$ denotes the radar cross-section (RCS) of the target. To emulate a challenging low-SNR scenario, we consider a small RCS of $0.01\text{ m}^2$ (typical for micro-drones), which results in SNRs from \mbox{-21.8~dB} at 40~m down to \mbox{-52.9~dB} at 240~m.
\begin{figure}[!t]
\centerline{\includegraphics[width=0.7\linewidth, height=10cm, keepaspectratio]{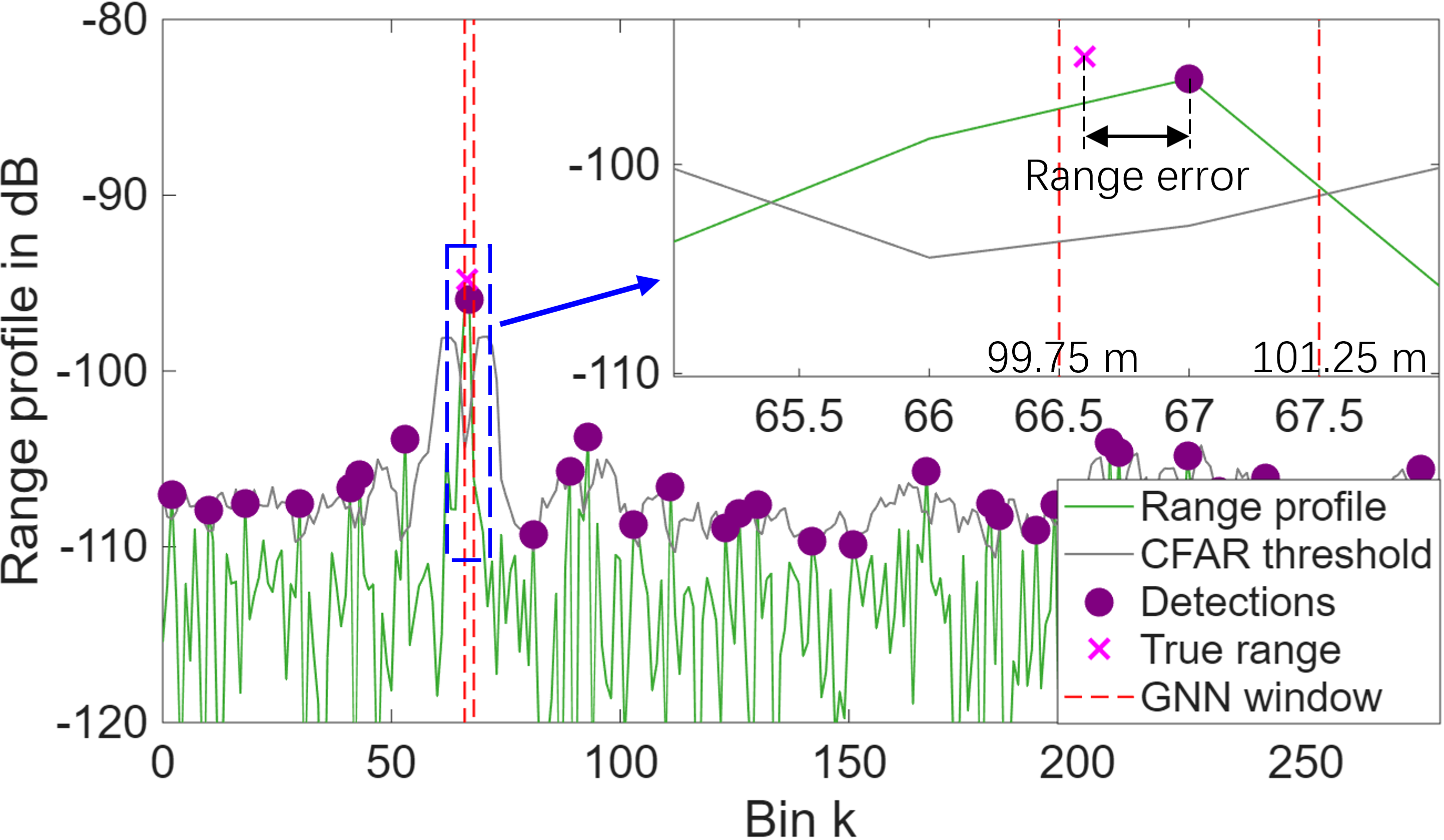}}
	\caption{Illustration of 1-bin detection window of GNN.}
	\label{fig_4}
\end{figure}

For a meaningful comparison, the CFAR thresholds are chosen according to the characteristics of each algorithm. The RP-based baseline (Pattern~2) uses a relatively strict $P_\text{fa}$ of $10^{-3}$ to balance missed detections and false alarms; as illustrated in Fig.~\ref{fig_3}(a), relaxing $P_\text{fa}$ (e.g., to $10^{-2}$) makes the periodogram output false-alarm dominated. In contrast, the proposed algorithm consists of CFAR followed by periodic-peak cross-pattern validation, which allows using a higher CFAR $P_\text{fa}$ of $10^{-1}$ to retain weak peaks while controlling the final false alarms via validation. Both approaches utilized a 1-bin detection window with global nearest neighbor (GNN) matching. The 1-bin window in GNN refers to a range span with a width of 1~bin. A target is considered detected when, after CFAR processing, at least one detection exists within this window surrounding the target's true range; the detection closest to the true range is used to calculate the range error. Otherwise, the event is recorded as a missed detection. For example, as illustrated in Fig.~\ref{fig_4}, a 1-bin window spanning 66.5--67.5 bins represents a range span of 99.75~m to 101.25~m. Additionally, a detection (either a potential target or a false alarm) is considered valid only when at least 4 out of the 6 theoretically expected bins (3 periodic bins in each of the two range profiles) show CFAR detections.
\begin{figure*}[t]
	\centering
	\subfloat[]{\label{fig_5a}\includegraphics[width=0.655\columnwidth]{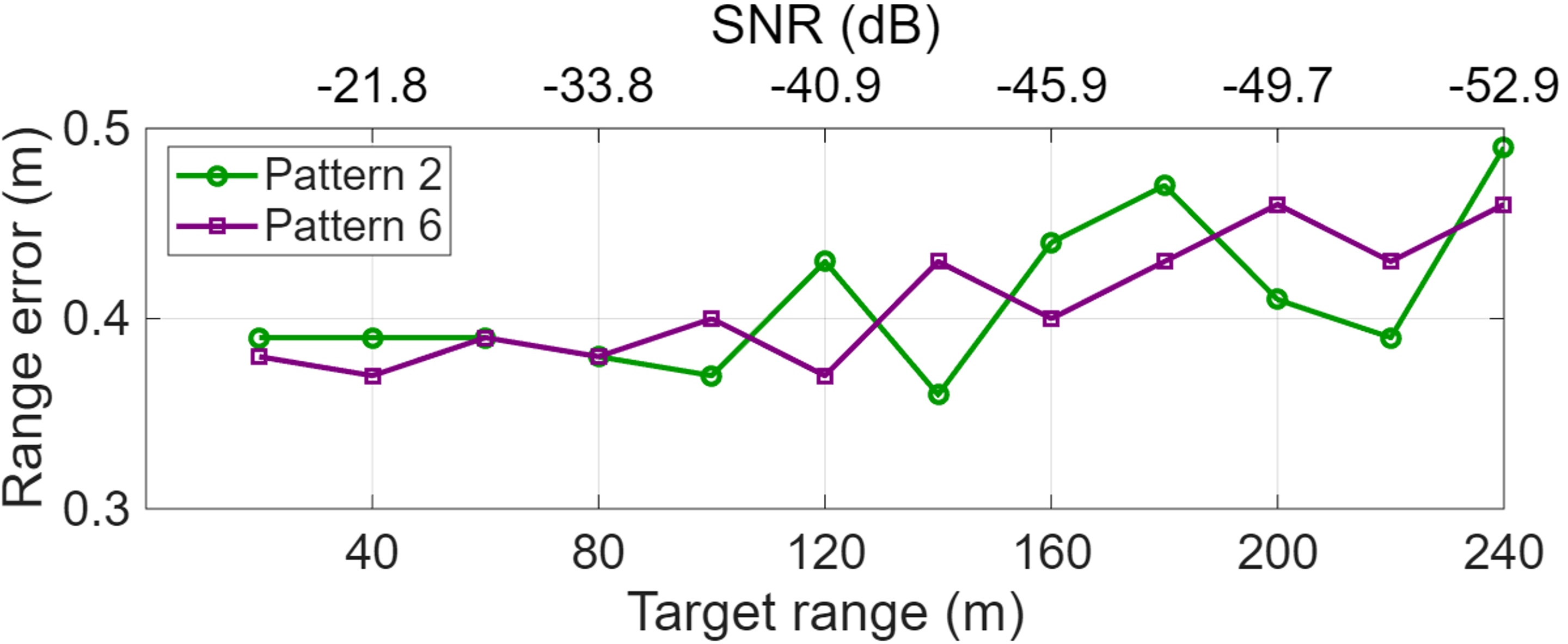}}\quad
   	   	\subfloat[]{\label{fig_5b}\includegraphics[width=0.655\columnwidth]{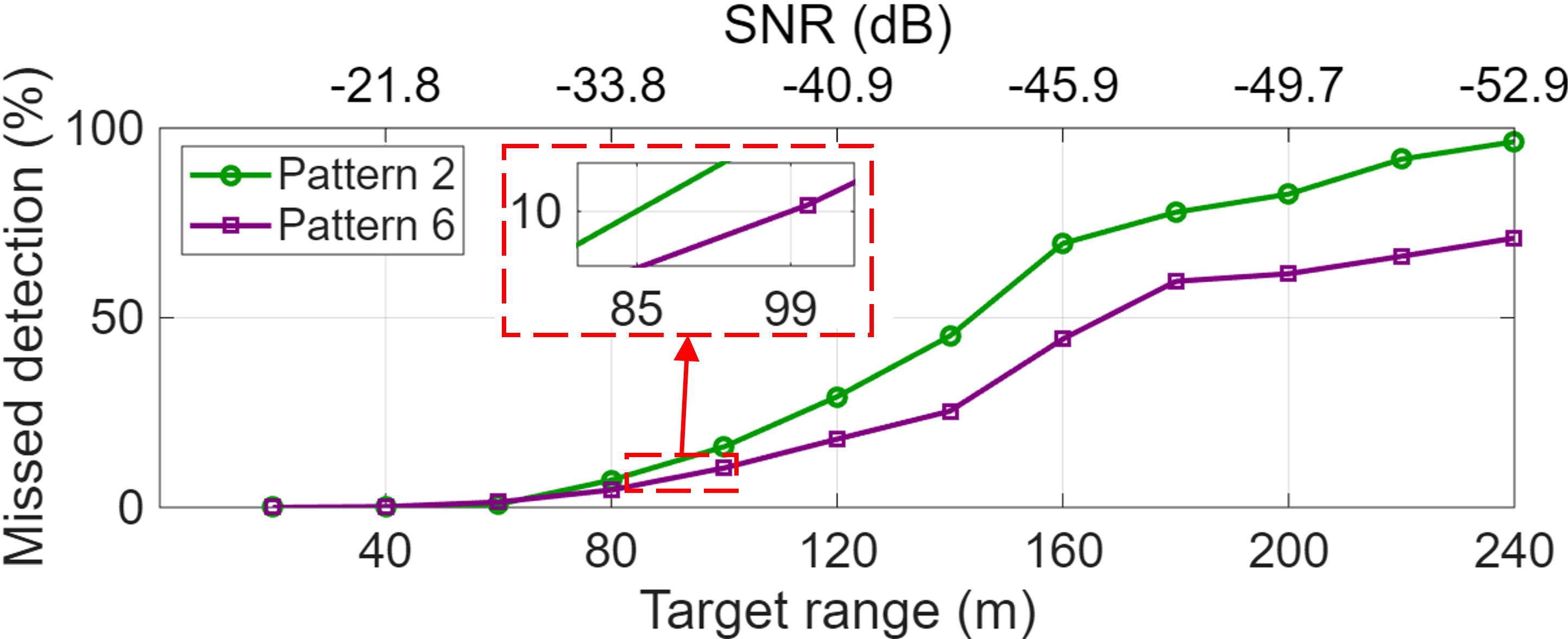}}\quad
        \subfloat[]{\label{fig_5c}\includegraphics[width=0.655\columnwidth]{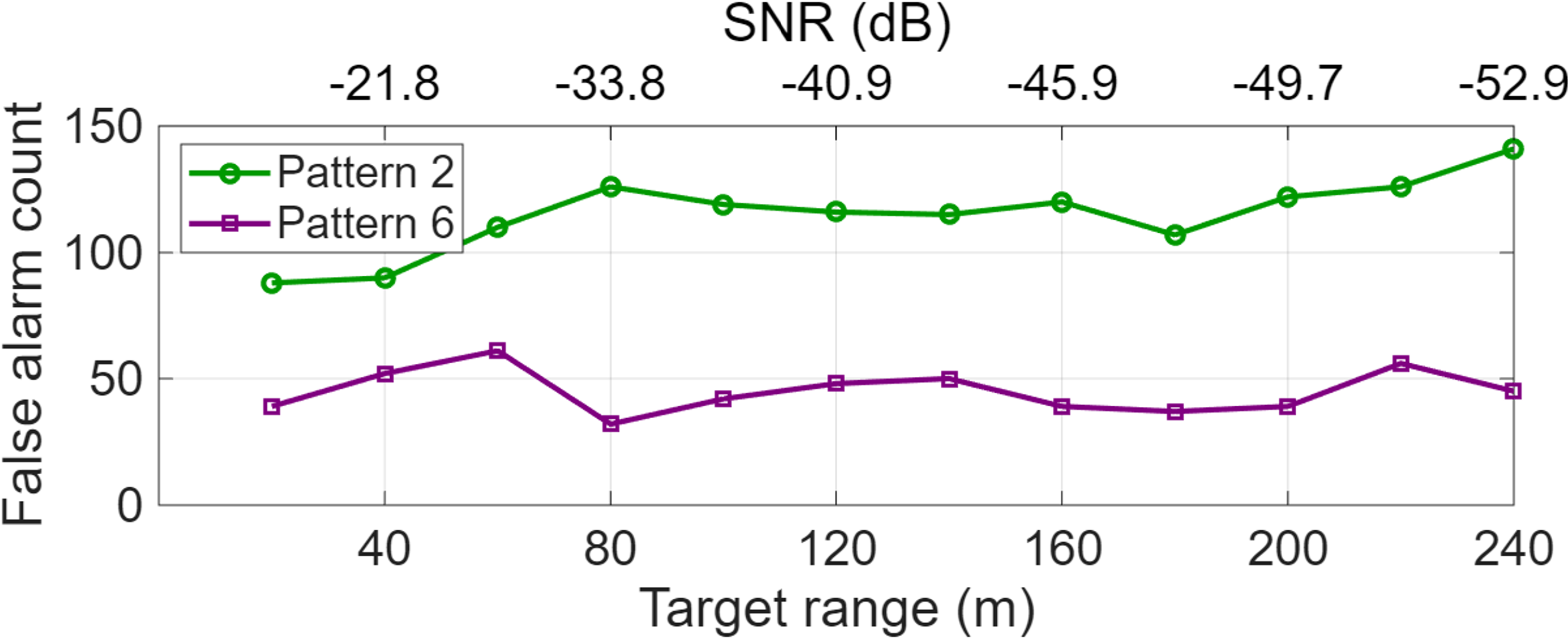}}\\
	\caption{Performance comparison between Pattern 2 and Pattern 6 across varying ranges (SNR): (a) range error, (b) missed detection rate, (c) false alarm count.}
	\label{fig_5}
\end{figure*}

Fig.~\ref{fig_5} presents the comparative performance results across multiple metrics. Fig.~\ref{fig_5}(a) reveals that both methods achieve comparable range error (approximately 0.4~m). This consistency is a direct consequence of the 1-bin GNN detection window, which ensures that whenever a target is successfully detected (i.e., when a detection exists within the window), the maximum range error is constrained to $\frac{\Delta R}{2}$. By employing this narrow GNN window across both methods, we establish a controlled environment where range accuracy remains consistent, allowing for a fair comparison of missed detection rate and false alarm.

As shown in Fig.~\ref{fig_5}(b), our approach demonstrates significantly lower missed detection rate. Particularly noteworthy is the performance at 10\% missed detection-a critical benchmark for radar systems. At this threshold, Pattern~2 achieves a maximum detection range of only 85~m, while Pattern~6 extends the reliable detection range to 99~m, representing a 16.5\% improvement in effective coverage.

Fig.~\ref{fig_5}(c) further highlights our method's superior false alarm suppression capability despite using a higher $P_\text{fa}$ value. Across all 6000 simulations, Pattern~2 generated 1380 false alarms, while Pattern~6 produced only 540, a 61\% reduction compared to the conventional approach. These performance improvements stem from the proposed periodic-peak consistency validation across two range profiles, which filters random noise-induced detections by requiring them to satisfy the expected peak periodicity and cross-profile agreement---a condition that random noise fluctuations rarely meet simultaneously.

\subsection{Enlarged Unambiguous Range}
The proposed algorithm generates range profiles with periodic peaks separated by $N_\text{g}$ bins. This periodicity provides a significant advantage: even when the peak corresponding to the true range falls outside the visible bin range, accurate range estimation remains possible. This characteristic substantially extends the unambiguous range compared to conventional periodogram algorithms. For our approach, the unambiguous range is determined by the least common multiple (LCM) of the group numbers from both patterns. With $N_\text{g,1}$ and $N_\text{g,2}$ denoting the numbers of groups in the respective patterns, the periodic peak locations repeat every $N_\text{g,1}$ and $N_\text{g,2}$ bins, and their joint repetition period is $\operatorname{lcm}(N_\text{g,1},N_\text{g,2})$ bins. Accordingly, the unambiguous range $R_\text{max}$ can be expressed as:
\begin{equation}\label{eqn_18}
R_\text{max} = \operatorname{lcm}(N_\text{g,1},N_\text{g,2})\Delta R.
\end{equation}

\begin{figure}[t]
	\centering
	\subfloat[]{\label{fig:h}\includegraphics[width=0.45\columnwidth]{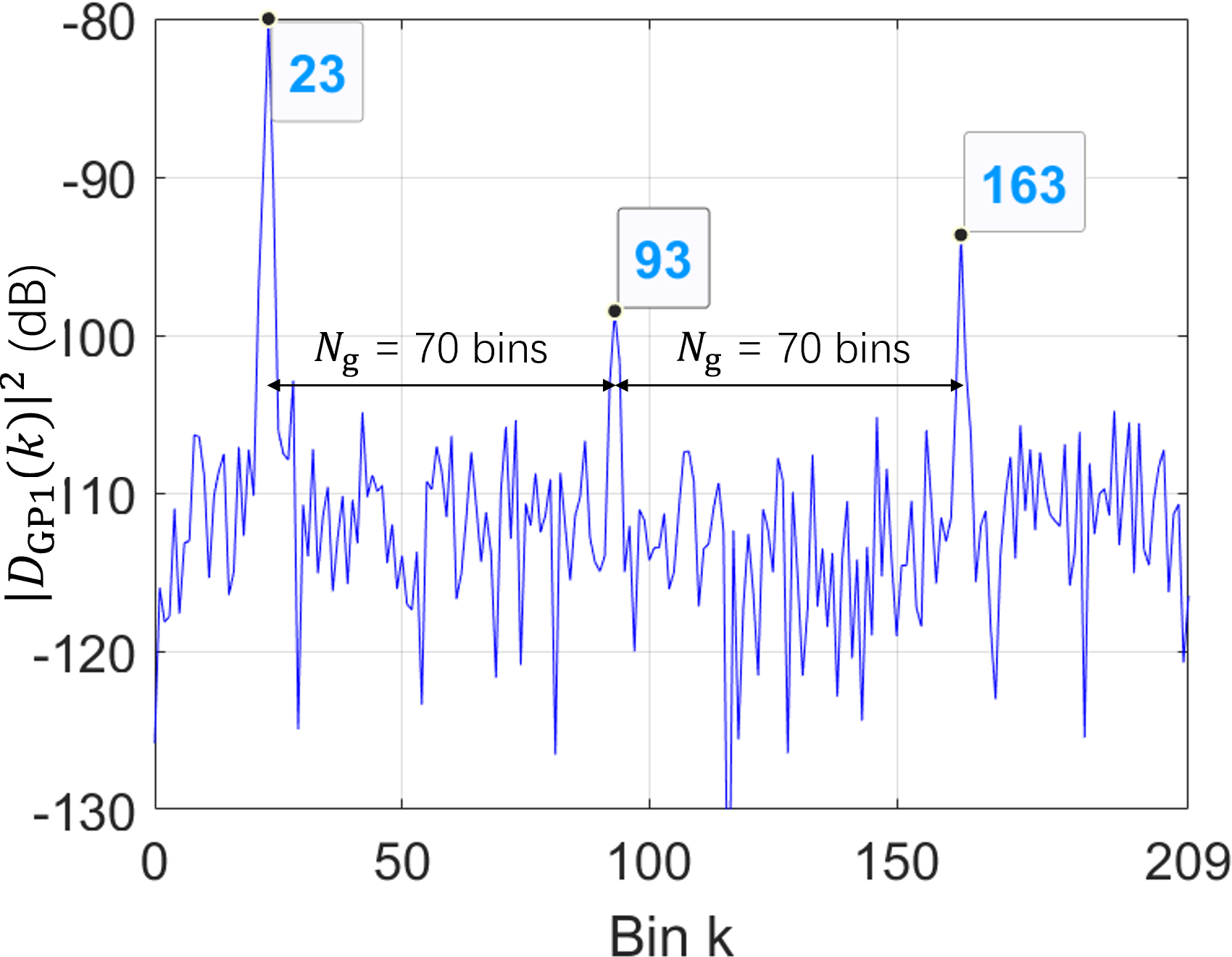}}\quad
   	   	\subfloat[]{\label{fig:i}\includegraphics[width=0.45\columnwidth]{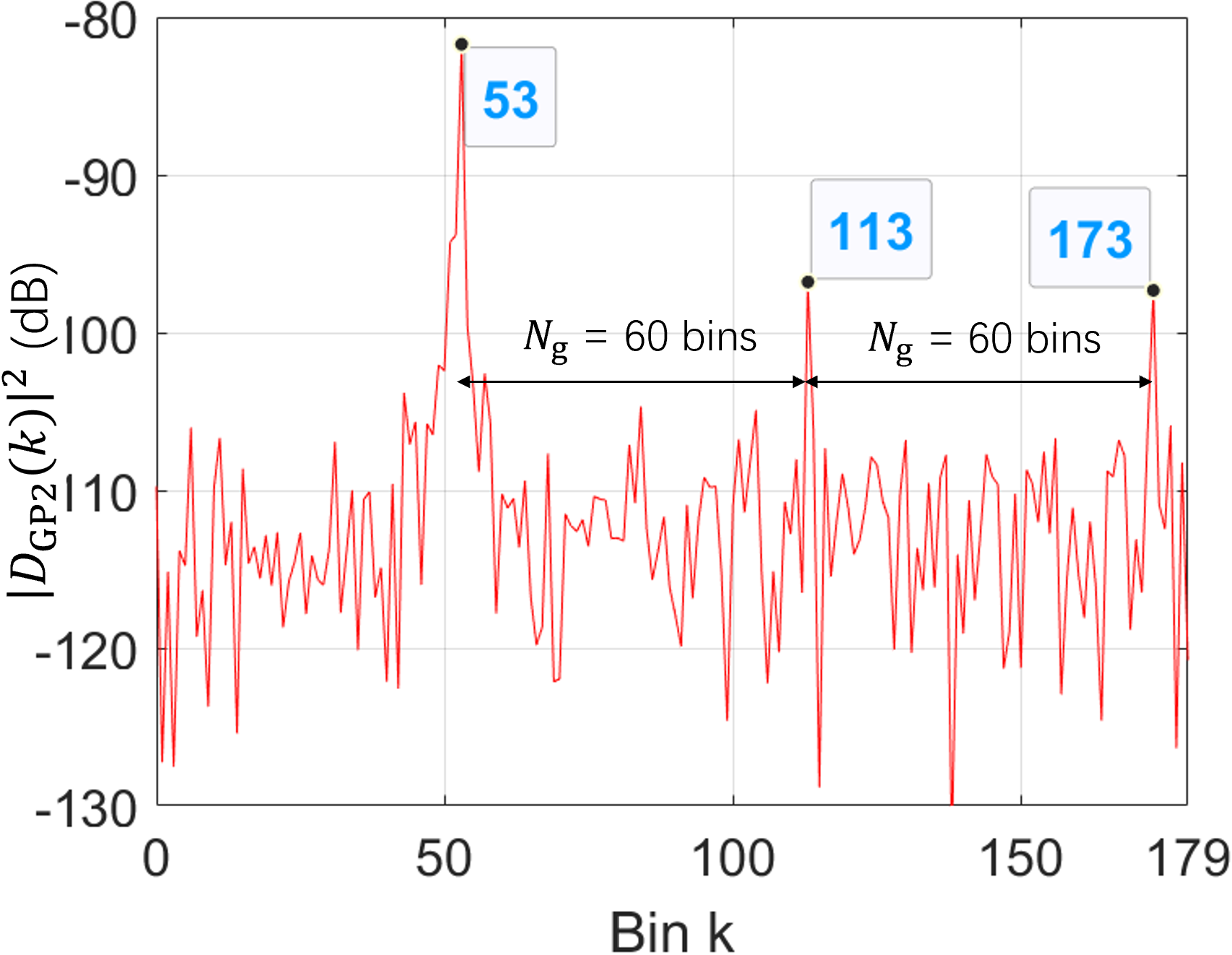}}\\
	\caption{Range profiles for a target at 350~m (beyond the visible bin span): (a) Pattern~3, (b) Pattern~4.}
	\label{fig_6}
\end{figure}

Figs.~\ref{fig_6}(a) and \ref{fig_6}(b) demonstrate this capability by presenting range profiles for a target at 350~m using Patterns~3 and~4, respectively. Within the visible bins, these profiles exhibit no common peaks. However, by leveraging the periodic nature of the peaks, we can infer the existence of a peak at bin~233 ($163+70$) in Fig.~\ref{fig_6}(a) and a peak at bin~233 ($173+60$) in Fig.~\ref{fig_6}(b). The inferred coincident peaks at the same bin provide compelling evidence of a target located at 349.5~m.

\subsection{Discussions}
The selection of parameters $N_\text{g}$ and $N_\text{s}$ follows strategic principles for optimizing performance. First, $N_\text{g}$ should be a divisor of $N_\text{sc}$ to avoid subcarrier fragmentation. Second, different patterns should employ non-multiple $N_\text{g}$ values to prevent false peak alignment across patterns. The sensing performance directly benefits from larger $N_\text{s}$, which increase processing gain at the cost of higher SS overhead.

The proposed algorithm introduces moderate computational complexity compared to conventional algorithms. While the periodogram algorithm requires $\mathcal{O}(N)$ complexity for $N$ bins, our method needs $\mathcal{O}(N)$ for each pattern plus $\mathcal{O}((k_1+k_2)N_\text{s})$ operations for periodic-peak consistency validation and false alarm suppression, where $k_1$ and $k_2$ represent detection counts in each range profile. This incremental complexity is well justified by the substantial performance improvements.

Despite using shorter FFTs (210-order for Pattern~3 and 180-order for Pattern~4 versus 280-order for Pattern~2), therefore lower processing gain, the proposed approach achieves superior performance by relying not only on peak amplitude but also on the distinctive periodic structure and periodic-peak consistency validation, as demonstrated in the false alarm filtering at bin~8 with extremely low peak amplitude in Fig.~\ref{fig_3}(c).

\section{Conclusions}\label{sectionV}
This paper introduces a GP and a multi-periodogram algorithm for range estimation in ISAC systems. By partitioning subcarriers into groups with an identical intra-group configuration, the proposed design generates range profiles with periodic peaks that can be factorized into a periodic component and an amplitude-shaping function. Two GPs can be jointly employed for cross-pattern peak validation to resolve the periodic-peak ambiguity and suppress noise-induced detections. Moreover, the final sensing configuration is obtained by superposing two GPs (as illustrated by Pattern~6 in Fig.~1), which reduces dedicated sensing overhead. Simulation results demonstrate improved performance over conventional approaches.


\bibliographystyle{IEEEtran}
\bibliography{ref}

\vfill

\end{document}